\documentclass[11pt]{article}
\usepackage{moriond,epsfig}

\def\Journal#1#2#3#4{{#1} {\bf #2}, #3 (#4)}

\def\plb{{\em Phys. Lett.}  B}
\def\prd{{\em Phys. Rev.} D}
\def\pr{{\em Physics Reports}}

\def\fun#1#2{\lower3.6pt\vbox{\baselineskip0pt\lineskip.9pt
        \ialign{$\mathsurround=0pt#1\hfill##\hfil$\crcr#2\crcr\sim\crcr}}}

\renewcommand\({\left(}
\renewcommand\){\right)}

\newcommand\eq[1]{Eq.~(\ref{#1})}

\newcommand\ee{\end{equation}}
\newcommand\be{\begin{equation}}
\newcommand\eea{\end{eqnarray}}
\newcommand\bea{\begin{eqnarray}}

\newcommand\GeV{\,\mbox{GeV}}

\newcommand\mpl{M_{\rm P}}

\newcommand\lsim{\mathrel{\rlap{\lower4pt\hbox{\hskip1pt$\sim$}}
    \raise1pt\hbox{$<$}}}
\newcommand\gsim{\mathrel{\rlap{\lower4pt\hbox{\hskip1pt$\sim$}}
    \raise1pt\hbox{$>$}}}

\newcommand\diff{\mbox d}

\def\call{{\cal L}}

\def\calp{{\cal P}}
\def\calr{{\cal R}}

\newcommand\bfx{{\bf x}}

\newcommand\sub[1]{_{\rm #1}}

\begin{document}
\vspace*{4cm}
\title{Generating the primordial curvature  perturbation from 
inflation\footnote{Invited talk at Rencontres de Moriond (The Cosmological
Model) March 2002, to appear in the proceedings}}

\author{David H.\ Lyth}

\address{Physics Department, Lancaster University, Lancaster LA1 4YB, U.K.}

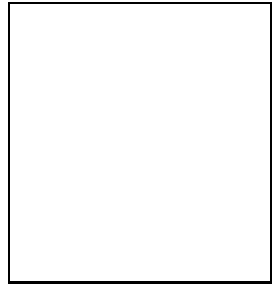
\maketitle\abstracts{The scale--independence of the
primordial curvature perturbation suggests that it comes from
the vacuum fluctuation during inflation of a light scalar field.
This field may be the inflaton, or a different `curvaton' field.
The observation of  primordial non--gaussianity 
would be a smoking gun for the curvaton model, while the 
observation of  gravitational waves originating during slow--roll inflation
would rule out the model.}

\section{Introduction}

Particle physics in the regime beyond the Standard Model is  certainly relevant
for the early Universe. As a result, astronomical observations of many 
different kinds have the potential to explore this regime.

The particular 
concern for this talk is the origin of the primordial curvature 
perturbation, which is now known to be the dominant cause of structure in the
Universe \cite{book}. 
At the outset I would like to draw attention to a paradigm shift,
which has taken place so slowly  that it has gone
largely unnoticed. Twenty years ago, when inflation was proposed, the received
wisdom was that practically all fields have gauge interactions.
 As a result
there was supposed to be 
a desert between the TeV and the GUT  energy scales, containing 
essentially no new physics. In particular, nothing significant was supposed
to  happen in the early Universe 
as the energy density falls from the GUT to the TeV scale. The paradigm shift
that I have in mind is the gradual population, actually starting in the late
seventies,  of the desert by
 gauge singlet fields to do important jobs, such as
\begin{itemize}
\item eliminate the CP--violating  $\theta$ term of QCD 
(axion and supersymmetric partners)
\item generate the $\mu$ term of the MSSM
\item generate neutrino masses (seesaw mechanism)
\item provide an origin for supersymmetry breaking (string moduli)
\item provide a lepto- or baryogenesis mechanism (eg.\ a late--decaying
right--handed sneutrino)
\item provide a dark matter candidate (wimpzilla)
\item implement inflation (the inflaton, which however need not be
a  gauge singlet)
\item {\em provide a `curvaton' field \cite{p01david}, 
which is not the inflaton but is 
nevertheless the origin of the primordial curvature perturbation}
(the focus of this talk).
\end{itemize}

It is gratifying that 
gauge singlets seem to be quite natural in the context of 
 string theory.
A frequent feature
 of cosmologies involving gauge singlets is a rather low 
final reheat temperature, corresponding to 
 the decay of a long--lived  gauge singlet which 
is not the inflaton. 

\section{The primordial curvature perturbation $\zeta$}

The curvature perturbation $\zeta$ is  defined \cite{bst,wmll}
through
the line element on spacetime slices of uniform energy density $\rho$,
\be
\diff\ell^2 = a^2(t) \delta_{ij} \diff x^i \diff x^j \(
1 + 2\zeta(\bfx,t) \)
\ee
It is of interest only on scales far outside the horizon, and on such scales
it is practically the same as the curvature perturbation $\calr$ defined
\cite{bardeen80,lyth85} on the slices orthogonal to comoving worldlines
(ie.\ such slices practically coincide with the uniform--density slices).
In contrast with the curvature perturbation on other slices
 (such as the Bardeen potential $\Phi$ defined on slices orthogonal to
zero--shear worldlines) $\zeta\simeq \calr$
 is time--independent  on super--horizon scales provided that the
pressure perturbation is adiabatic \cite{bardeen80,wmll}. 
Here, `adiabatic'  means that
the pressure $P$ is uniform  on the uniform--energy--density slices, or in 
other words that  there is a unique relation $\rho(P)$ throughout spacetime.
This holds  during complete matter--domination ($P=0$) and
during complete radiation--domination ($P=\rho/3$) and also during 
single--field inflation. It may fail though when there is a mixture of
matter and radiation, the extreme case being the curvaton model that I shall
discuss.

In the present  context `primordial' denotes the epoch in the early Universe 
a few Hubble times  {\em before} scales of cosmological interest start to
enter the horizon. At this epoch the Universe is supposed to be 
completely radiation--dominated so that the
 curvature perturbation $\zeta$ is time--independent. Observation is presently
{\em consistent} with the  hypothesis that $\zeta$ is (i) gaussian
(eg.\ the non--linear parameter $f\sub{NL}$ is $\lsim 10^{3}$)
and (ii) has a  completely scale--independent spectrum (spectral 
index $n\simeq 1.0\pm 0.1$). How do these findings compare with theory?
On the usual hypothesis  that  $\zeta$ is generated by the inflaton
the answer is well--known; almost perfect gaussianity is predicted
(in particular $f\sub{NL}\sim 1$) but in most models one expects
significant scale--dependence allowing even present observation to discriminate
between models \cite{treview}. 
As we shall see the situation is almost reversed on
the  alternative `curvaton' hypothesis; significant non--gaussianity is
generic so that even the present observation constrains the parameter space,
while significant scale--dependence, though possible,  seems less likely than
under the usual hypothesis.

\section{Nature points us towards exponential inflation with four--dimensional
field theory}

The existence of a scale--independent primordial curvature perturbation
is beautifully explained by  the following rather conservative scenario
\cite{book,p01david}.
The observable Universe starts out inside the horizon (Hubble distance)
 during an  era of almost exponential inflation. As each cosmological scale
leaves the horizon, the vacuum fluctuation of some light scalar field
is converted to a classical scale--independent perturbation.
 This perturbation  then generates the curvature perturbation
either directly (the inflaton hypothesis) or indirectly (the curvaton 
hypothesis).

In this  scenario, one has to be able to integrate out any 
extra dimensions to give an effective  four--dimensional
 (4D) field theory. String theory and branes may play a crucial role, but only
indirectly in determining the form of the effective field theory.

An extraordinarily well-publicized alternative to inflation is the 
(now cyclic) ekpyrotic scenario \cite{ek}. Inflation is replaced by
collapse,  under the influence of an exponential potential $V(\phi)$ where
$\phi$ is measures the distance between a pair of colliding branes in one
extra dimension. In contrast with the case of inflation, the 
perturbation in $\phi$ generated by the vacuum fluctuation is strongly 
dependent on the slicing of spacetime (gauge) \cite{myek1}
and is scale--independent
only for a particular choice corresponding to the Bardeen potential $\Phi$. 
It is conjectured that the bounce will convert $\Phi$ to $\zeta$, but 
that seems to me  implausible because the bounce is described by
 stringy higher--dimensional physics  that does not `know' about 
4D  spacetime. Instead, I would expect the bounce
(if and when the necessary physics is defined)  to generate
perturbations in 4D spacetime 
that have nothing to do with pre--bounce conditions,
and do not lead to the required scale--independent primordial curvature
perturbation $\zeta$. By the same token I would expect the same to be true
in any other bounce scenario, including `pre--big--bang' \cite{pbb}
which invokes
stringy fields with no potential.

\section{Slow--roll inflation}
\label{secslowroll}

The simplest way of generating almost exponential inflation is through
the slow--roll mechanism \cite{book}. This invokes Einstein gravity, and a 
canonically--normalized inflaton field $\phi$ whose potential $V(\phi)$
satisfies
the flatness conditions $\epsilon\ll 1$ and $|\eta|\ll 1$ where
\bea
\epsilon &\equiv & \frac12 \mpl^2 (V'/V)^2 \\
\eta\equiv \mpl^2 V''/V &=& 3 V''/H^2
\,.
\eea
Because of the flatness conditions, the equations of motion
\bea
3\mpl^2 H^2(t) &=& V(\phi) + \frac12\dot\phi^2 \\
0 &=& \ddot\phi + 3H\dot\phi + V'
\eea
generally
 have the attractor solution $3H\dot\phi\simeq -V'$ leading as required
to $|\dot H/H^2|\ll 1$.

It is usually  assumed that the  4D  field theory containing
$\phi$ remains valid
after inflation. In that case the theory has to contain the Standard
Model, 
corresponding  to a lagrangian $\call(\phi,\cdots,{\rm SM\, fields})$. 
As a result  the flatness of the 
inflationary potential has to be protected against the effects
of  supergravity and radiative corrections. In particular,
the condition  $|V''|\ll H^2$ requires either (i) 
a cancellation in tree--level 
supergravity potential (implying a special form for the Kahler potential and
 superpotential) or (ii)  unsuppressed inflaton couplings which flatten the 
renormalization group improved potential leading to a `running mass' inflation
model.
\cite{treview}.

A radical alternative has recently been proposed \cite{dt}
(see \cite{kl02} for references to later work)
in which the 
transition from inflation to the conventional big bang (reheating) 
 involves stringy higher--dimensional physics
that {\em cannot} be described by 4D  field theory.
The inflaton $\phi$ is a stringy field (for instance the distance between
a pair of colliding branes) and the lagrangian during inflation 
contains only $\phi$ and maybe a few other stringy fields,
$\call(\phi,\cdots)$. The inflaton 
potential, determined directly by string theory, is supposed  to be
flat enough for inflation.
(In the original proposal \cite{dt} it was the same up to a constant as the
exponential inter--brane potential
invoked later for the ekpyrotic scenario.) It seems to me that 
 this `brane inflation' shares  the problem of ekpyrotic and other bouncing
scenarios;  the era immediately  preceding the conventional
big bang (in this case  the reheating era) does not `know' about 
4D spacetime,  and will presumably generate unviable
  4D perturbations
which  have nothing to do with the earlier era when scales are leaving
the horizon (in this case the era of inflation). 

\section{The inflaton scenario for generating the curvature perturbation}

During almost--exponential inflation every light field acquires an
almost scale--independent perturbation. Which of these fields is responsible
for the primordial curvature perturbation?
The almost universally accepted answer is; the inflaton! Indeed, the
inflaton automatically generates a curvature perturbation. A few
Hubble times after cosmological scales leave the horizon the spectrum
 and spectral index are 
  given by \cite{book}
\bea
\frac4{25}\calp_\zeta(k)
 &=& \frac1{75\pi^2\mpl^6}\frac{V^3}{V'^2} 
\label{delh} \\
n(k) -1 &\equiv & \frac{\diff \ln \calp_\zeta}{\diff \ln k}
= 2\eta-6\epsilon
\,,
\label{nofv}
\eea
where on each scale the 
 potential and its derivatives are
 evaluated  at the epoch of horizon exit
$k=aH$. This epoch is  given in terms of the field value at the end
of slow--roll inflation by
\be
\ln(k\sub{end}/k)\equiv N(k)
=\mpl^{-2}\int^\phi_{\phi\sub{end}} (V/V') \diff\phi
\,.\label{Nofv}
\ee
On cosmological scales $N\lsim 60$, the  value depending
on the history of the Universe  after inflation.

In `single--field' models of inflation, which are the most typical,
 the inflationary trajectory in field space is essentially unique,
and it determines the epoch at which  inflation gives way to
 matter-- or radiation--domination. The curvature perturbation in such models
is time--independent until the end of inflation. In `two--field' 
models  the flatness conditions
are satisfied in two  different field directions, and there is a 
family of possible inflationary 
 trajectories (lines of steepest descent) which are curved
in field space. In those models the spectrum of the
curvature perturbation increases somewhat
during inflation \cite{treview}.
In both   cases though, the value of $\zeta$
at the end of inflation is maintained
until cosmological scales start to enter the horizon, unless there is a 
curvaton field with the properties that I shall describe.
According to the inflaton hypothesis there is no curvaton field.
The prediction of the inflaton hypothesis may be written
\be
\calp_\zeta^\frac12({\rm inflaton}) = 
\calp_\zeta^\frac12({\rm observed})
\( \frac{XV^\frac14}{\epsilon^\frac14 \times 2\times 10^{16}\GeV}
\)^2
\label{delh2}
\ee
where $X=1$ in a single--field model and $X>1$ in a two--field (or
multi--field) model. In order for this prediction to agree with observation,
the bracket should be equal to 1.
Together with \eq{nofv} this
places a strong constraint on the inflationary potential,  ruling out or 
disfavoring several  attractive models. Such models usually regain their
attraction if they are   liberated from  the constraint by 
 the  curvaton hypothesis \cite{dl}.

\section{The curvaton scenario}

\subsection{The sequence of events}

Recently, attention \cite{p01david}
has been drawn to the
alternative possibility, that the primordial curvature perturbation
$\zeta$  is
generated by a `curvaton' field different from the inflaton.
The curvaton generates the primordial curvature perturbation through
the following sequence of events.\footnote
{The basic mechanism  was discovered by Mollerach
\cite{silvia}, and the possible  non--gaussianity of $\zeta$ 
 was noted by Linde and Mukhanov \cite{lm}.}

1. The  curvature perturbation $\zeta$ 
at the end of inflation is much less than the observed value. 
(The  case where it is comparable to that value without being the same,
 corresponding to a mixture of the 
inflaton and curvaton mechanisms, has not been considered so far and
seems contrived.)
 Requiring that it be less than say $1\%$
of the observed value, we learn from \eq{delh2} that
$V^\frac14 < 2\times 10^{15}\GeV$. With such a low inflation scale,
gravitational waves from inflation cannot have a detectable effect
on the cmb anisotropy \cite{gw02}! This is an anti--smoking gun:
{\em  assuming slow--roll inflation, a detection of gravitational
 waves in the cmb anisotropy would invalidate the pure curvaton model}.

2. The curvaton field $\sigma$ 
 acquires a  gaussian perturbation $\delta\sigma\ll \sigma$
with a nearly flat spectrum
given by $\calp_\sigma^\frac12\simeq  H/2\pi$. The flat spectrum
 certainly requires a flatness condition
$|\eta_{\sigma\sigma}|\ll 1$
where
\be
\eta_{\sigma\sigma} \equiv \mpl^2  \(\partial^2 V/\partial \sigma^2\) /V
\,.
\label{etasigsig}
\ee

3. The curvaton  and its fractional perturbation  survive
(the latter possibly attenuated \cite{dllr} by some factor $f$)
 until $H$ falls below the curvaton mass $m_\sigma$, 
which occurs during a radiation--dominated
era. The curvaton then starts to
oscillate at each point in space with amplitude
$\sigma+\delta\sigma(\bfx)$, corresponding to curvaton energy density
\be
\rho_\sigma(\bfx)=\frac12 m_\sigma^2 \(\sigma+ \delta\sigma(\bfx)\)^2
\simeq \frac12 m_\sigma^2 \sigma^2 \( 1+ 2\frac{\delta\sigma(\bfx)}{\sigma} \)
\,.
\ee
 At this stage there is still no curvature
perturbation but there is an approximately gaussian isocurvature density
perturbation, specified by the gauge--independent  entropy perturbation
\bea
S_{\sigma,{\scriptstyle \rm rad}}& \equiv&
\frac13 \frac{\delta\rho_\sigma}{\rho_\sigma} -
\frac14 \frac{\delta\rho\sub{rad}}{\rho\sub{rad}}\\
&\simeq & \frac13 \frac{\delta\rho_\sigma}{\rho_\sigma}
\simeq \frac23  \frac{\delta\sigma}{\sigma}
\label{S}
\,.
\eea
where  `rad' denotes the radiation. In the second line,
 $\delta\rho_\sigma$ and $\delta\sigma$ are evaluated on flat slices.

4. The curvaton has no gauge coupling, and any  Yukawa couplings are
 suppressed.
As a result it  survives  for some time, during which 
$r\equiv \rho_\sigma/\rho\sub{rad}$ grows like the scale factor $a$.
During this time $S_{\sigma,{\scriptstyle \rm  rad}} $ 
remains constant but a curvature perturbation  develops
given by
\be
\zeta(t) = \frac{3r(t)}{4+3 r(t)} S_{\sigma,{ \scriptstyle \rm rad}} 
\label{curvpred1}
\ee

5. The curvaton decays into radiation, after which $\zeta$ is constant.
The prediction for the spectrum of the
 primordial density perturbation is therefore
\be
\calp_\zeta^\frac12 \simeq 
\frac f\pi \left. \frac{r}{4+3r}  \right| \sub{decay}
\left. \frac H\sigma \right| \sub{inflation}
\label{curvpred}
\,,
\ee
For a successful prediction the right hand side must be equal to 
$5\times 10^{-5}$.

In the curvaton model, the spectral index defining the
scale--dependence is
 given in terms of the derivatives of the potential at horizon exit
 by \cite{p01david,luw}
\be
n-1 = 2\eta_{\sigma\sigma} - 2\epsilon
\,,
\ee

This may  be compared with the usual inflaton scenario  which gives
$n-1=2\eta-6\epsilon$. 

\subsection{A new way of handling  curvature and isocurvature
perturbations}

Let me pause to give a simple derivation of the prediction \eq{curvpred1},
based on a new way \cite{p01david,wmll} of handling the perturbations.
The starting point as usual is the assumption that we have
 $N$  fluids
each with 
a definite equation of state $P_i(\rho_i)$. The total energy density
is
\be
\rho = \sum_i \rho_i
\ee
and each fluid separately satisfies the continuity equation
\be
\dot\rho_i = -3H (\rho_i+ P_i)
\ee
The new idea is to work with the $N$  curvature  perturbations
$\zeta_i$, defined on slices of uniform $\rho_i$. They are given
in terms of the separate density perturbations (defined on flat slices)
by the standard expression
\be
\zeta_i =  H \Delta t_i = H\frac{\delta\rho_i}{\dot\rho_i}
=-\frac13 \frac{\delta\rho_i}{\rho_i + P_i}
\ee
where $\Delta t_i$ is the displacement between  flat-- and uniform--density
slices. 
The separate curvature perturbations are time--independent on super--horizon
scales \cite{wmll}, and so are the entropy perturbations 
\be
S_{ij} \equiv \zeta_i - \zeta_j
\ee
but the total curvature perturbation is time--dependent,
\be
\zeta(t)=\frac{\sum_i \dot\rho_i(t) \zeta_i}{\rho(t)}
\ee
In our case $N=2$, the fluids being
the curvaton with $P=0$ and the radiation with $P=\frac13\rho$,
which leads immediately to \eq{curvpred1}.

\subsection{Non--gaussianity}

In the final equality of \eq{S} the quadratic term
 term has been neglected. Including it gives
\be
\zeta \simeq  \zeta\sub{gauss} + f\sub{NL}
\( \zeta\sub{gauss}^2 - \overline{ \zeta\sub{gauss}^2 } \)
\ee
where $\zeta\sub{gauss}\sim 10^{-5}$ 
is the gaussian contribution whose spectrum is
given by \eq{curvpred}
 and
\be
f\sub{NL} \sim \left. \frac\rho{\rho_\sigma} \right|\sub{decay}
\sim 10^5 \left. f \frac H{\sigma} \right|\sub{inflation}
\ee
At present the observational  bound 
is $f\sub{NL}\lsim 10^3$ which already
constrains the parameters of the curvaton model. In contrast, the inflaton
hypothesis gives $f\sub{NL}\sim 1$ which is practically unobservable. 
{\em A future detection
$f\sub{NL}\gg 1$ would be a smoking gun for the curvaton hypothesis.}

Another   possible signature of the curvaton model, which I shall not discuss
here, is an isocurvature perturbation in the baryon or neutrino
density that is  {\em directly related to the curvature perturbation}
because it is generated by the curvaton decay \cite{luw}. Such an isocurvature
perturbation would  automatically be of
 potentially {\em observable}
magnitude and would be  {\em $100\%$ correlated} with the curvature
perturbation. This is in contrast with the usual mechanism for 
producing an  isocurvature perturbation (CDM, baryon or neutrino), namely to
invoke a field different from  the one producing the curvature
perturbation. With such a  mechanism  there is no reason why the isocurvature
perturbation should be of observable magnitude,  and also  it is typically 
uncorrelated with the curvature perturbation. Partial correlation occurs in 
two--field models \cite{langlois}, but only in  the very special case that
the field causing the isocurvature 
perturbation is a combination of the two fields (the one
 orthogonal
to the inflaton at the end of inflation.)

\subsection{Who is the curvaton?}

As with the inflaton, the
flatness condition $|\eta_{\sigma\sigma}|\ll 1$ during inflation is
difficult to achieve in the context of supergravity. More seriously,
the flatness condition will probably have to be preserved after inflation
until $H$ falls below the true mass $m_\sigma$, to avoid wiping out
\cite{dllr} the
perturbation in $\sigma$. The latter requirement especially suggests that
the curvaton may be  a pseudo--goldstone boson, whose potential is under all
circumstances kept flat by a global symmetry. In that case one might
expect $\eta_{\sigma\sigma}$ to be tiny,  typically implying 
a spectral index very close to 1.

Nevertheless, the attractive
alternative that the curvaton may correspond to a flat direction of
global supersymmetry is worth investigating \cite{dllr}. Candidates in the
latter case might be a right--handed sneutrino \cite{my} or a string
modulus \cite{moroi}.

\subsection{Conclusion}

The only known mechanism for generating the observed
 scale--independent  primordial curvature perturbation is the traditional
one, involving
 exponential inflation  in the context of  a four--dimensional field
theory which is an extension of the Standard Model of particle physics.
In this scenario, 
extra dimensions and branes may play an important role in determining the
form of the field theory, but they are not directly relevant.
The scenario does {\em not} require that the inflaton is responsible
for the curvature perturbation. Some other  `curvaton' field might
do the job, and its presence might be signalled by
primordial non--gaussianity.

\end{document}